\documentclass[twocolumn,showpacs,superscriptaddress]{revtex4}

\usepackage{amsmath}
\usepackage{amssymb}
\usepackage{graphicx}
\usepackage{dcolumn}
\usepackage{bm}
\usepackage{subfigure}
\usepackage{color}
\usepackage{ifpdf}
\usepackage{epstopdf}
\usepackage{CJK}
\usepackage[
  pdfstartview=FitH,%
  bookmarks=true,%
  bookmarksopen=true,%
  breaklinks=true,%
  colorlinks=true,%
  linkcolor=blue,anchorcolor=blue,%
  citecolor=blue,filecolor=blue,%
  menucolor=blue,pagecolor=blue,%
  urlcolor=blue]{hyperref}

\begin{document}

\begin{CJK}{GBK}{song}
\title{The observation of diffraction phases in matter wave scattering}
\author{Xuguang Yue}
\affiliation{School of Electronics Engineering $\&$ Computer Science, Peking University, Beijing 100871, China}
\author{Yueyang Zhai}
\affiliation{School of Electronics Engineering $\&$ Computer Science, Peking University, Beijing 100871, China}
\author{Zhongkai Wang}
\affiliation{School of Electronics Engineering $\&$ Computer Science, Peking University, Beijing 100871, China}
\author{Hongwei Xiong}
\affiliation{Department of Applied Physics, Zhejiang University of Technology, Hangzhou 310023, China}
\author{Xuzong Chen}
\affiliation{School of Electronics Engineering $\&$ Computer Science, Peking University, Beijing 100871, China}
\author{Xiaoji Zhou}
\thanks{Electronic address: xjzhou@pku.edu.cn}
\affiliation{School of Electronics Engineering $\&$ Computer Science, Peking University, Beijing 100871, China}
\date{\today}

\begin{abstract}
We study the diffraction phase of different orders via the Dyson expansion series, for ultracold atomic gases scattered by a standing-wave pulse. As these diffraction phases are not observable in a single pulse scattering process, a temporal Talbot-Lau interferometer consisting of two standing-wave pulses is demonstrated experimentally with a Bose-Einstein condensate to explore this physical effect. The role of the diffraction phases is clearly shown by the second standing-wave pulse in the relative population of different momentum states. Our experiments demonstrate obvious effects beyond the Raman-Nath method, while agree well with our theory by including the diffraction phases. In particular, the observed asymmetry in the dependence of the relative population on the interval between two standing-wave pulses reflects the diffraction phase differences. The role of interatomic interaction in the Talbot-Lau interferometer is also discussed.
\end{abstract}

\maketitle

\section{Introduction}
\label{sec:intro}

The phase of a wave function plays an interesting role in quantum mechanics. It plays a special and basic role in the quantum behaviour of a system. On the one hand, the introduction of an extra phase for a whole quantum state does not change the expectation value of an observable quantity. On the other hand, the phase difference between two quantum states may lead to observable interference effect. Atom interferometry, one of the most promising technologies suggested by modern atomic physics, is based on the phase differences between two or more states via corresponding paths~\cite{Cronin2009rmp}. In an atom interferometer, the Kapitza-Dirac (KD) scattering of matter waves from a standing wave is a very powerful and versatile tool for the coherent splitting and/or mixing of momentum modes~\cite{Meystre2001book, Gupta2001review}.

The study on the KD scattering of atoms by light has been numerous, both in experiments~\cite{Deng1999PRL, Hagley1999prl, Ovchinnikov1999PRL, Sapiro2009njp, Gupta2002prl, Xiong2011pra, Li2008prl, Huckans2009pra, Campbell2005prl, keller1999apb} and in theories~\cite{Muller2008pra, Buchner2003pra, Edwards2010pra, Trippenbach2000jpb}. According to the intensity and duration of the atom-light interaction, it is usually divided into the Bragg, the Raman-Nath and the channeling regimes~\cite{keller1999apb}. The Bragg scattering, which is limited to regime of low intensity and long pulse, relies severely on the energy-momentum conservation (Bragg condition)~\cite{Meystre2001book}. In the high intensity and short pulse, the so-called Raman-Nath (RN) regime~\cite{Gupta2001review, Meystre2001book, Cronin2009rmp}, the atomic displacement along the standing wave is small which leads to the neglecting of the atomic kinetic energy term in the Hamiltonian. In the channeling regime, numerical methods are usually needed~\cite{keller1999apb, Ovchinnikov1999PRL, Xiong2011pra}. Nevertheless, there is still lack of an analytical expression for the diffraction amplitude over the whole regime, not even an approximate one. Besides, the diffraction phase is rarely studied. In Ref.~\cite{Henkel1994jp2}, Henkel \emph{et al.} discussed the diffraction phase ``near" the RN regime by expanding the action integral along the classical trajectory of atoms perturbatively.

Here we study the diffraction of an atom from a standing-wave pulse by an evolution operator method~\cite{Sakurai}, which is not limited by the intensity or duration of the pulse, compared with the usual method for Bragg scattering, RN regime and channeling regime. The evolution operator is calculated through the Dyson expansion series. The elements of the evolution matrix are found to be complex, giving rise to nonzero diffraction phases. Applying the evolution operator to an atom initially in the zero momentum state, approximate analytical expressions for the diffraction amplitudes of different orders are obtained, from which the corresponding diffraction phases can be calculated straightforwardly. The analytical dependence of the amplitudes and phases on the pulse intensity and duration shows good consistency with the numerical results.

The diffraction phase has no observable effect in the relative population in the one-pulse scattering. Interferometry is one of the methods to observe the diffraction phase~\cite{Cronin2009rmp}. However, it is not easy to detect the above mentioned diffraction phase in interferometers realized in the RN regime or with the Bragg scattering, partially because the diffraction phase is small in the RN regime or it may cancel in some interferometers.

In order to explore the effect of the diffraction phases of different orders,  a temporal Talbot-Lau (TL) interferometer~\cite{Deng1999PRL, Gerlich2007np, Xiong2013arxiv, Sapiro2009pra} is demonstrated experimentally with a Bose-Einstein condensate (BEC). The temporal TL interferometer consists of two identical KD pulses separated by an interval, which is varied in the experiment. The intensity and duration of each pulse are large enough, so that the diffraction is beyond the RN regime, as a result of which the amplitude of the second diffraction order is considerable. The relative population of the BEC around the zero momentum state after the TL interferometer is measured as a function of the interval. In the population oscillation, an asymmetry induced by the diffraction phase differences is observed between the rising and the falling edges. The experimental observation agrees well with our analysis including the high-order diffraction phases. We further discuss the role of interatomic interaction in the TL interferometer.

The paper is organized as follows. In Sec. II, we calculate the evolution operator for an atom scattered by a standing-wave pulse and apply it to an atom initially in the zero momentum state. In Sec. III, the effect of the diffraction phases is studied experimentally and theoretically using a temporal TL interferometer with a BEC. A brief summary and discussion is given in Sec. IV.

\begin{figure}[t]
\centering\includegraphics[width=8cm]{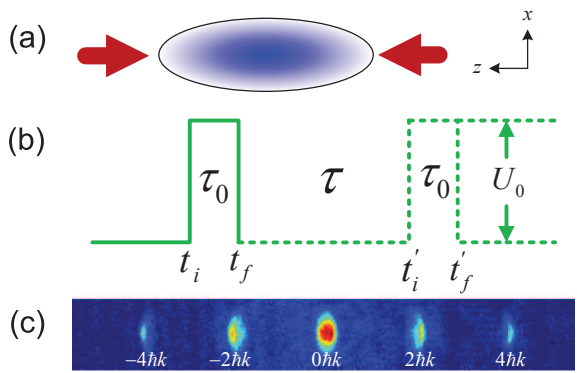}
\caption{(Color online) (a) Scheme of the KD scattering of a BEC along the $z$ direction. (b) Time sequences of one pulse and two pulses. The one pulse process, which is illustrated by solid curves, is turned on between $t_{i}$ and $t_{f}$. The strength and duration of the pulse is $U_{0}$ and $\tau _{0}$, respectively. A second identical pulse is applied after a variable interval $\tau $ (dotted curve). (c) A typical TOF image obtained in our experiment. As we can see, the fraction of $\pm4\hbar k$ order is considerable.}
\label{fig:scheme}
\end{figure}

\section{The scattering amplitude and phase for KD scattering}
\label{sec:theory}
\subsection{The evolution operator}

We first consider the diffraction of an atom by a standing-wave pulse as shown in Fig.~\ref{fig:scheme}. The standing wave pulse consists of a pair of linearly polarized,
counter-propagating laser beams along the $z$ direction (Fig.~\ref{fig:scheme}(a)). The duration of the pulse, which is turned on at $t_i$ and turned off at $t_f$, is $\tau_0=t_f-t_i$ (Fig.~\ref{fig:scheme}(b)). The detuning of the laser beams is far enough \text{so that} we can ignore the spontaneous emission.

The Hamiltonian for a single atom diffracted by the standing wave light field is given by
\begin{equation}
    \hat{H}=\frac{\hat{p}^{2}}{2M}+f(t)V(\hat{z}),
    \label{eq:hamiltonian}
\end{equation}
where the first term in the right hand side is the single particle kinetic Hamiltonian $\hat{H}_0$ with the atomic momentum operator $\hat{p}$ and atomic mass $M$. The second term in the right hand side is the time-dependent scattering potential, with $f(t)$ describing the shape of the pulse. $V(\hat{z})=U_{0}\cos^{2}(k\hat{z})$ represents the effective interaction between the atom and the light field, which is a spatially periodic potential~\cite{Denschlag2002jpb, Morsch2006}, with the position operator $\hat{z}$, the trap depth $U_{0}$ and the wave vector $k=2\pi/\lambda$ ($\lambda$ is the wavelength of the laser).

Due to the position operator $\hat{z}$, $e^{\pm ik\hat{z}}$ can be described as a momentum shift operator: $e^{\pm ik\hat{z}}|p\rangle=|p\pm\hbar k\rangle$~\cite{Meystre2001book}. In the momentum representation, the periodical potential $V(\hat{z})$, apart from a constant term, can be written in an operator form as $\hat V=\frac{U_{0}}{4}\int \mathrm{d}p(|p+2\hbar k\rangle\langle p| +|p-2\hbar k\rangle\langle p|)$, which explicitly shows the momentum change $2\hbar k$ of an atom scattered by the standing wave~\cite{Gadway2009OE}. This process is accomplished by absorbing a photon from one of the laser beams and emitting it into the counter-propagating one. The iteration of such a two-photon process results in an equally spaced distribution separated by $2\hbar k$ in the momentum space.

With a standard procedure~\cite{Sakurai}, the evolution operator for the one-pulse scattering process can be obtained via the Dyson expansion series:
\begin{equation}
U(t_{f},t_{i})=\sum_{n=0}^{\infty}U^{(n)}(t_{f},t_{i}),
\label{eq:UtExpansion}
\end{equation}
where the zeroth order term $U^{(0)}(t_{f},t_{i})=1$ is the unitary operator, and
\begin{equation}
\begin{split}
U^{(n)} =&\left(\frac{-i}{\hbar}\right)^{n}%
\int_{t_{i}}^{t_{f}}\mathrm{d}t_{1} \cdots
\int_{t_{i}}^{t_{n-1}}\mathrm{d}t_{n} [V_{I}(t_{1})\cdots V_{I}(t_{n})]
\end{split}
\label{eq:Un}
\end{equation}
is the $n^{th}$ order term with $t_{i}\leq t_{n}\leq t_{n-1}\leq\cdots\leq t_{1}\leq t_f$. Here $V_{I}(t)=f(t)e^{i\hat{H}_0t/\hbar}\hat{V}e^{-i\hat{H}_{0}t/\hbar}$ is the Hamiltonian in the interaction picture. Each $V_I$ in Eq.~(\ref{eq:Un}) stands for a single two-photon scattering process, and their product at different time represents all possible configuration of the $n^{th}$ order two-photon processes. With the expression for $\hat{V}$, it is straightforward to get
\begin{equation}\label{eq:VI}
    \begin{split}
    V_{I}(t) =&\frac{U_{0}}{4}f(t)\int \mathrm{d}p\left(e^{i\delta_{+}(p)t}|p+2\hbar k\rangle\langle p|\right. \\
    &+\left.e^{i\delta_{-}(p)t}|p-2\hbar k\rangle\langle p|\right),
\end{split}
\end{equation}
where $\delta_{\pm}(p)=4\omega_{R}\pm2kp/M$ is the detuning of the two-photon process for an atom of momentum $p$, which simply reflects the energy and momentum conservation for a free particle. The second term of $\delta_\pm(p)$ is the Doppler shift~\cite{Stenger1999prl}, which is neglected in Ref.~\cite{Henkel1994jp2}.

The evolution operator $U(t_f,t_i)$ can be expressed as a matrix of infinite dimension, whose matrix element
\begin{equation}
a_{l,m}(q;t_f,t_i)=\langle q+2l\hbar k|U(t_f,t_i)|q+2m\hbar k\rangle
\label{eq:alm}
\end{equation}
represents the transition amplitude from the initial state $|q+2m\hbar k\rangle$ to the final state $|q+2l\hbar k\rangle$ in the interaction picture, with $-\hbar k \leq q\leq \hbar k$ the quasi-momentum. From Eq.~(\ref{eq:alm}) we can see that, $a_{l,m}$ depends not only on the quasi-momentum $q$, but also on the starting time $t_i$ and the ending time $t_f$ of the pulse. Additionally, some properties of $a_{l,m}$ can be drawn out without knowing its explicit expression. Firstly, the evolution matrix $U(t_f,t_i)$ is symmetric with respect to the main diagonal, which means $a_{l,m}=a_{m,l}$, because of the real Hermite operator $\hat{V}$. Secondly, the transition probability should not depend on the starting time but the duration of the pulse, so one has $|a_{l,m}(q;t_f,t_i)| =|a_{l,m}(q;\tau_0)|$.

The matrix element $a_{l,m}(q;t_f,t_i)$ can be expanded as,
\begin{equation}
\begin{split}
a_{l,m}=a^{(0)}_{l,m}+a^{(1)}_{l,m} +a^{(2)}_{l,m}+\cdots
\end{split}
\end{equation}
where $a^{(n)}_{l,m}$ is the $n^{th}$ order expansion coming from the corresponding term in the Dyson series. Substituting Eq.~(\ref{eq:VI}) into Eq.~(\ref{eq:Un}), we have
\begin{subequations}
\begin{align}
a^{(0)}_{l,m}&=\delta_{l,m}, \label{eq:alm:0}\\
a^{(1)}_{l,m}&=-\frac{iU_0}{4\hbar}\int_{t_i}^{t_f} \mathrm{d}t
f(t) e^{i\delta_\pm(p)t}\delta_{l,m\pm1}, \label{eq:alm:1}\\
a^{(2)}_{l,m}&=\left(-\frac{iU_0}{4\hbar}\right)^2 \sum_\ell \int_{t_i}^{t_f} \mathrm{d}t\int_{t_i}^t  \mathrm{d}t_1  \label{eq:alm:2} \\
& f(t) f(t_1) e^{i\delta_\pm(p)t}e^{i\delta_\pm(p)t_1}\delta_{l,\ell\pm1}\delta_{\ell,m\pm1}.\notag
\end{align}
\label{eq:an_expansion}
\end{subequations}
The transition amplitude of the zeroth order approximation (\ref{eq:alm:0}) is independent of $\tau$ and equals unity only when the momentum state does not change, otherwise it equals zero. Eq. (\ref{eq:alm:1}) and Eq. (\ref{eq:alm:2}) reflect the first and second order of the interaction between the atom and the laser pulse, respectively. The Kronecker $\delta_{l,m}$ function is due to the special form Eq.~(\ref{eq:VI}) of the periodical scattering potential, reflecting the momentum change $2\hbar k$ in a single two-photon process. However, cascade of two-photon processes may result in no momentum change, which can be seen by the expansion of the two Kronecker $\delta_{l,m}$ functions.

As we have mentioned, $a_{l,m}$ depends on both $t_i$ and $t_f$, which comes from the kinetic operator $\hat{H}_0$. The dependence can be removed through the picture transformation by introducing the transition amplitude in the Schr{\"o}dinger's picture $\tilde{a}_{l,m}= \exp[-iE_l(q)t_f/\hbar] {a}_{l,m}\exp[iE_m(q)t_i/\hbar]$ ~\cite{Sakurai}. Here $E_l(q)$ and $E_m(q)$ are eigen-values of $\hat{H}_0$ for states $|2l\hbar k+q\rangle$ and $|2m\hbar k+q\rangle$, respectively. Then $\tilde{a}_{l,m}$ depends only on the quasi-momentum $q$ and the pulse duration $\tau_0$.

It is noticed that the scattering amplitude $\tilde{a}_{l,m}$ is a complex quantity, characterized by a modulus and a phase
\begin{equation}
    \tilde a_{l,m}(q;t_f,t_i)=|\tilde{a}_{l,m}(q;\tau_0)|e^{i\phi_{l,m}(q;\tau_0)}.
\end{equation}
The modulus determines the transition probability that the atom is scattered from state $|2m\hbar k+q\rangle$ to state $|2l\hbar k+q\rangle$. The phase $\phi_{l,m}(q;\tau_0)$, which stems from the detuning $\delta_{\pm}(p)$ as we can see from Eq.~(\ref{eq:an_expansion}), is known as the diffraction phase~\cite{Buchner2003pra}. The study on the diffraction phase is rare, even it is of crucial importance in the atom interferometry, as it may modify the interference signals.


The diffraction phase has no observable effect in the relative population in the one-pulse scattering, as it does not change the transition probability, which is what we can measure after the pulse when all orders are well separated. However, it plays a special role in the interference of two-pulse scattering. As we will show in Sec.~\ref{sec:exp}, if another pulse is applied after some variable interval $\tau$, the role of the diffraction phase can be clearly seen.


\subsection{Diffraction amplitude and phase for an atom}

As a special case, we consider the diffraction of an atom by a square shaped standing-wave pulse as shown by the solid line in Fig.~\ref{fig:scheme}(b). In this case, the initial momentum $p=0$, which means $m=0$, $q=0$, and $f(t)=\theta(t-t_i)- \theta(t_f-t)$ with $\theta(t)$ the Heaviside step function. We denotes $\tilde{a}_{l,m=0}(q=0;t_f,t_i)$ and $\phi_{l,m=0}(q=0;\tau_0)$ by $a_l(\tau_0)$ and $\phi_l(\tau_0)$, respectively. As $q=0$, $\delta_+=\delta_-$, the diffraction should be symmetric with respect to the zeroth order, which means $a_l=a_{-l}$. Substituting the value of $m$ and $q$ and the expression of $f(t)$ into Eq.~(\ref{eq:an_expansion}), up to the second order, we can get approximate expressions for the scattering amplitude $a_l$, which depend on the pulse depth $s=U_0/E_R$ in unit of the single photon recoil energy $E_{R}=\hbar\omega_{R} =\hbar^{2}k^{2}/2M$ and duration $\tau_0$ explicitly:
\begin{subequations}
\begin{align}
a_{0} & =1-\frac{s^{2}}{128}\left(1-it_0-e^{-it_0}\right), \label{eq:phase:0}\\
a_{\pm1} & =\frac{s}{16}\left(1-e^{-it_0}\right), \label{eq:phase:1}\\
a_{\pm2} & =\frac{s^{2}}{3072}\left(1-4e^{3it_0}+3e^{4it_0}\right), \label{eq:phase:2}
\end{align}
\label{eq:phase}
\end{subequations}
with $t_0=4\omega_R\tau_0$, which is a basis unit of the scattering process. The diffraction phase obtained from $e^{i\phi_l}=a_l/|a_l|$ for different order can be extracted directly from the corresponding expressions. When $t_0\ll1$, $\phi_l$ reduces to $|l|\pi/2$.

\begin{figure}[t]
\centering\includegraphics[width=9cm]{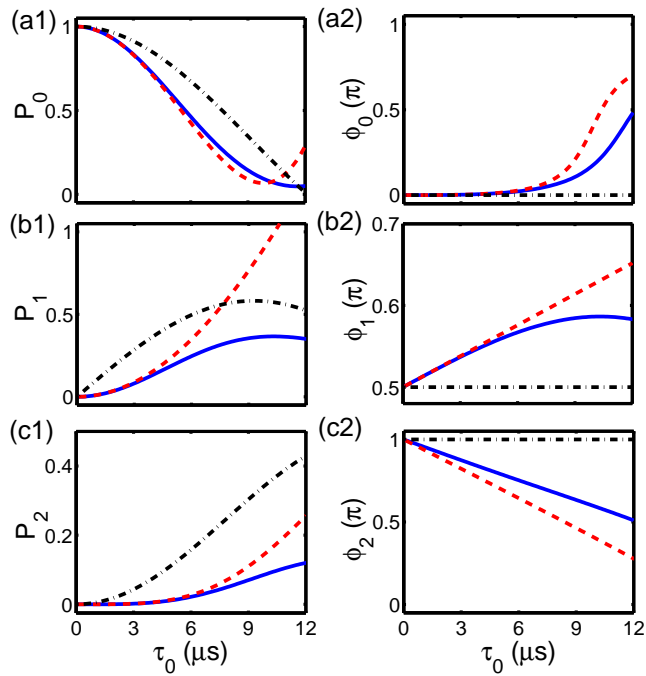}
\caption{(Color online) (a1)-(c1) the population $P_l$ and (a2)-(c2) the diffraction phase $\phi_l$ of the order $l=0,1,2$ vs the pulse duration $\tau _{0}$, respectively. The strength of the pulse is $s=20$. The blue solid, red dashed and black dash-dotted lines represent results of Num, Eq.~(\ref{eq:phase}) and RN approximation, respectively. }
\label{fig:phase}
\end{figure}

Speaking of the KD scattering, one has to mention the well-known RN approximation, which neglects the effect of kinetic energy during the pulse and is quite often used in the field of atom optics~\cite{Meystre2001book, Gupta2001review}. The transition amplitude of the RN approximation is simply $c_l=i^nJ_l(\Theta)$~\cite{Sapiro2009njp, Deng1999PRL}, with $\Theta=U_0\tau_0/2\hbar$ and $J_l(x)$ the Bessel function of the first kind. The scattering phase in RN approximation is thus $\phi^\prime_l=|l|\pi/2$, where the absolute value of $l$ comes from the fact $J_{-l}=(-1)^lJ_l$. This is consistent with the results of Eq.~(\ref{eq:phase}).

In Fig.~\ref{fig:phase}, we show the dependence of the population $P_l=|a_l|^2$ and the diffraction phase $\phi_l$ (l=0,1,2) on the pulse duration $\tau_0$ with $s=20$, respectively, according to Eq.~(\ref{eq:phase}) (red dashed line). Results by exact diagonalization of the Hamiltonian (\ref{eq:hamiltonian}) numerically~\cite{Xiong2011pra} (for convenience we refer to this method as Num hereafter) (blue solid line) and RN approximation (black dash-dotted line) are shown together for comparison. As shown by Fig.~\ref{fig:phase}(a1)-(c1), the populations given by all methods are consistent with each other when the duration $\tau_0$ is small. These three methods deviate as the pulse duration increases. Nevertheless, the trends of the amplitudes from different models are almost the same over a large range of pulse duration. However, the dependences of the diffraction phases on the pulse duration (Fig.~\ref{fig:phase}(a2)-(c2)) disagree qualitatively. Although all methods give the same result of $|l|\pi/2$ when $\tau_0 \rightarrow 0$, the phase $\phi_l$ given by Eq.~(\ref{eq:phase}) and Num varies drastically as $\tau_0$ increases, while that given by RN approximation keeps a constant of $|l|\pi/2$. The difference comes from the inclusion of the kinetic energy in the former two methods. Moreover, the approximate results of both population and phase by Eq.~(\ref{eq:phase}) agree excellently with the numerical result when $\tau_0<10\mu$s, which shows the effectiveness of our expansion method. The failure of the approximate results at long duration is because of the finite expansion order and can be corrected by expansion to higher orders.


\section{The effect of the diffraction phases in the TL interferometer}
\label{sec:exp}

In Sec.~\ref{sec:theory}, we have shown that the diffraction phases of different orders are not zero. In the one-pulse scenario, it is difficult to measure the effect of the diffraction phase. In this section, we show the effect of the diffraction phase using a TL interferometer with a BEC.

We first describe our experimental setup and procedure. The experiment was performed using an elongated ${}^{87}$Rb BEC created in a magnetic trap as in our previous experiments~\cite{Xiong2011pra, Zhou2010pra, Yang2008pra}. The condensate, containing $N_0=1\times10^5$ atoms in the internal state $5~^2{S_{1/2}},\left| {F = 2,{m_F} = 2}\right\rangle$, has a Thomas-Fermi radius of $8 (80) \mu$m in the radial (axial) direction. The axial and radial frequencies of the magnetic trap are $20$~Hz and $220$~Hz, respectively. The BEC in the magnetic trap was illuminated with an optical standing wave created by a retroreflected $\pi$-polarized laser beam along the axial direction~(see Fig.~\ref{fig:scheme}(a)). Losses in the retroreflected beam were negligible. The laser at the wavelength $\lambda=852$~nm was focused to a waist of $110~\mu$m to cover the condensate. After a variable time $\tau$, a second identical pulse was applied (Fig.~\ref{fig:scheme}(b). Then the light pulse and the magnetic trap were shut off simultaneously. The momentum distribution of the condensate was imaged after $28$~ms free expansion, long enough for different momentum modes to be resolved (Fig.~\ref{fig:scheme}(c)). The atomic number of each momentum state was determined by integrating the optical density of the absorption image and denoted as $N_l$. The relative population of the momentum state $|{2l\hbar k}\rangle $ can be extracted by $P_l=N_l/\sum N_l$.

\begin{figure}[t]
\centering \includegraphics[width=8.5cm]{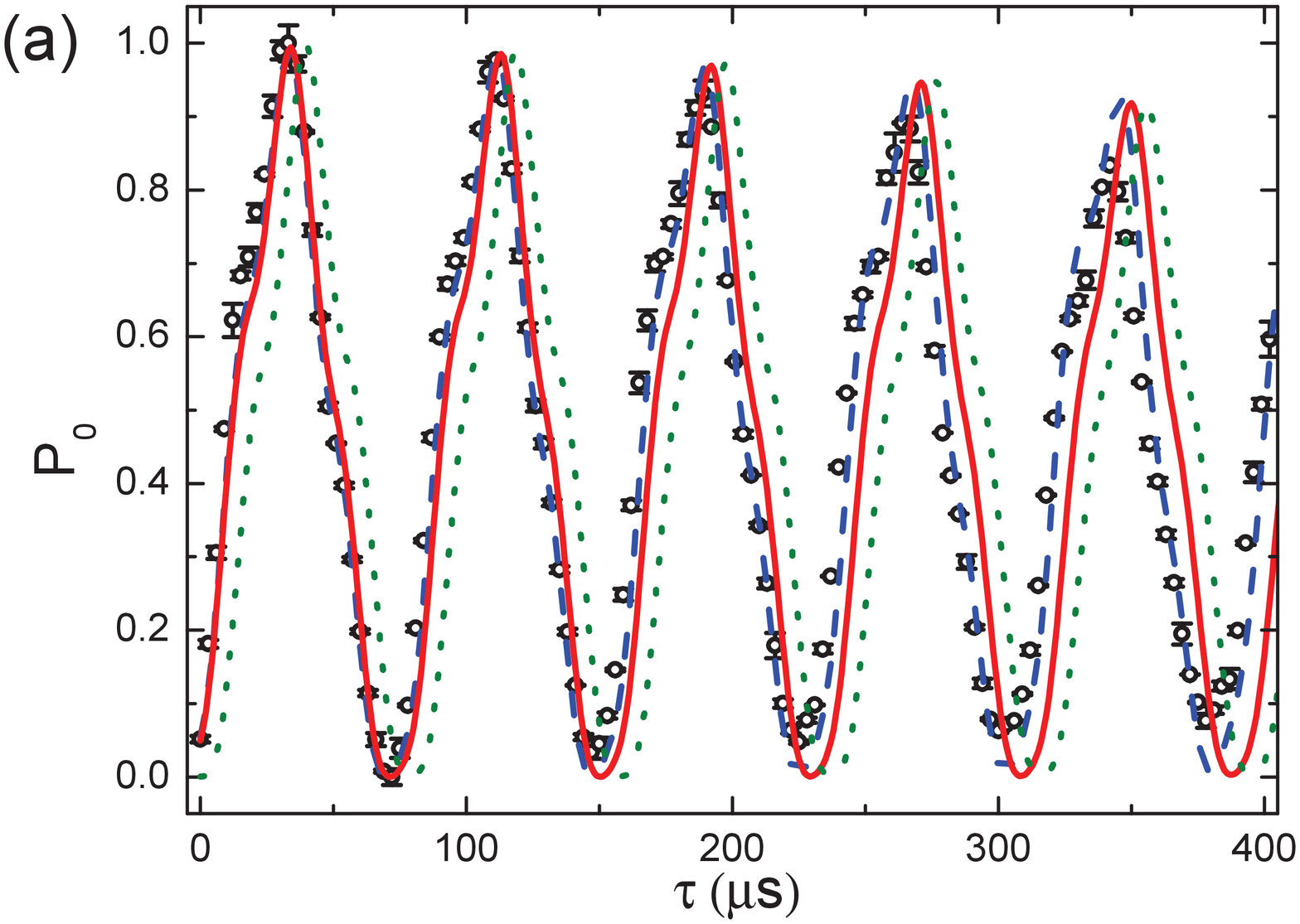}\\
\centering \includegraphics[width=8.5cm]{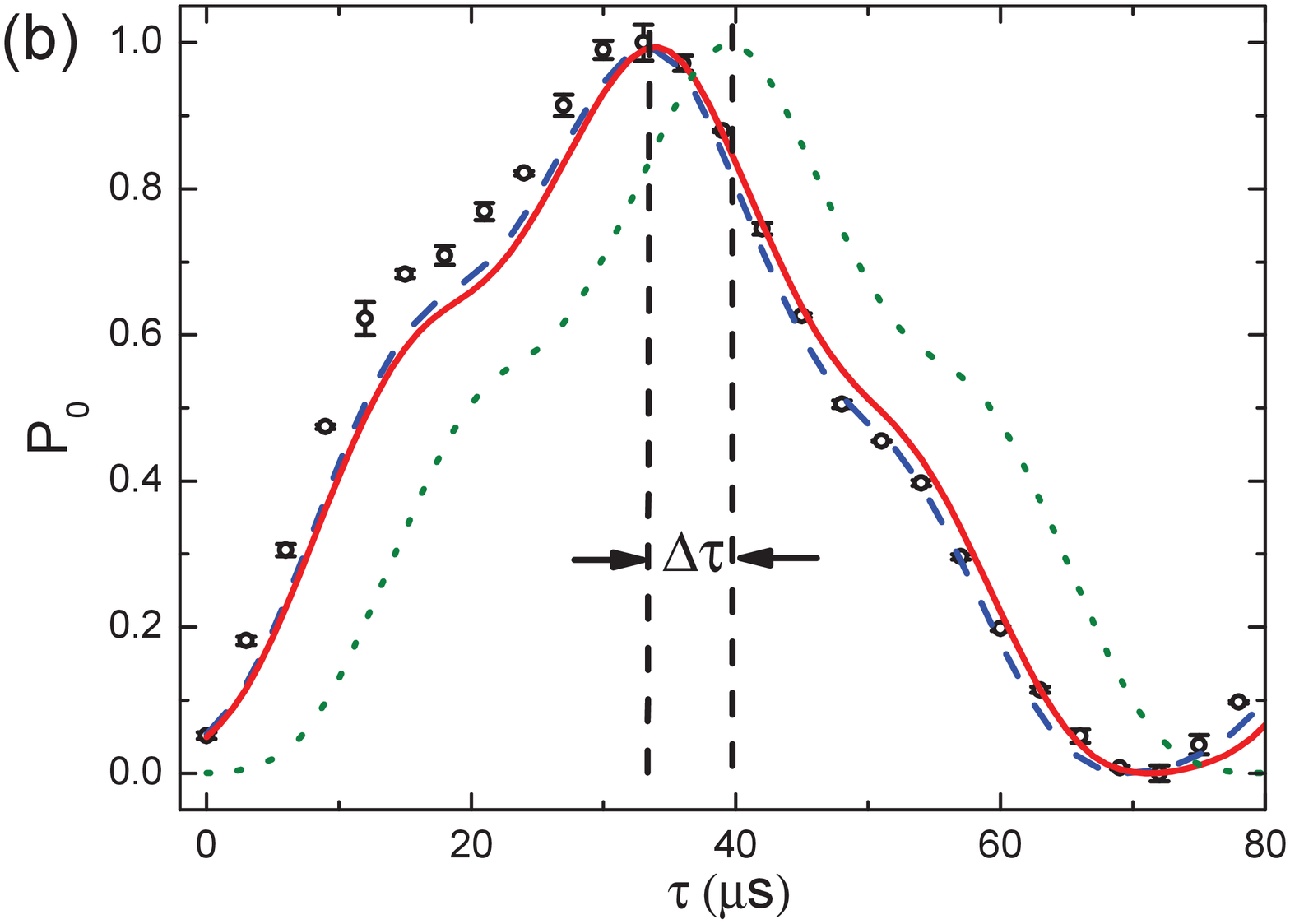}
\caption{(Color online) (a) The normalized relative population $P_{0}$ of $0\hbar k$ as a function of the interval $\tau $. The error bars correspond to the standard deviation of three consecutive measurements of the same $\tau$. Numerical results by Eq.~(\ref{eq:2pls}) (red solid line),  GP equation (blue dashed line) and RN approximation (oliver dotted line) including finite momentum spread are shown for comparison. (b) Details of (a) within the first Talbot period. The two vertical dashed lines indicate the position of maximum of Eq.~(\ref{eq:2pls}) and RN approximation, respectively, with distance $\Delta\tau$.}
\label{fig:mainresult}
\end{figure}

In the experiments, the depth of the pulse, which is calibrated by KD scattering, is set to $s=20$ and the durations of both pulses are $\tau_0=6~\mu$s. The intensity and duration of each TL pulse are large enough that the diffraction is beyond the RN regime (the RN regime is usually limited to $\tau_0 \ll 1/\sqrt{s}\omega_R \simeq 11~\mu$s ~\cite{Meystre2001book, Cronin2009rmp}), as a result of which the amplitude of the second order diffraction is considerable. The interval $\tau$ between the two pulses is changed every $3~\mu$s up to $400~\mu$s. In Fig.~\ref{fig:mainresult}(a), the relative population $P_0$ as a function of the varied interval $\tau$ is shown by the black circles. The error bars, corresponding to the standard deviation of three consecutive measurements of the same $\tau$, are very small and reflect the high repeatability of our experiments. It is clearly shown that $P_0$ oscillates with $\tau$ periodically, gaining its maximum and minimum near odd times of $39.5~\mu$s and multiples of $79~\mu$s, respectively. The oscillation amplitude damps as $\tau$ increases. Besides, there is an obvious asymmetry between the rising and falling edges of the oscillation in a period. In the rising edge there is an obvious turning point, which divides the rising edge into two parts, whose slopes differ apparently. In contrast, the slope of the falling edge changes little. To see the asymmetry more clearly, we show the detail of the first period of the oscillation in Fig.~\ref{fig:mainresult}(b).

The process of the TL interferometer can be formulated by the evolution operator developed in Sec.~\ref{sec:theory}. Assuming the starting and ending points of the second pulse as $t_{i}^\prime$ and $t_{f}^\prime$ (see Fig.~\ref{fig:scheme}(b)), respectively, the evolution operator of the second pulse is $U(t_{f}^\prime,t_{i}^\prime)$. The two-pulse scattering process can be then described by $|\Psi_2(t=t_f^\prime)\rangle =U(t_{f}^\prime,t_{i}^\prime)U(t_{f},t_{i})| \Psi(t=t_i)\rangle $ since the evolution operator during the interval equals unity in the interaction picture. With the matrix elements $a_{l,m}$ of evolution operator $U$, we can get the population $P_m$ of the momentum state $|2m\hbar k+q\rangle$ after the two-pulse sequence: $P_m(q,\tau_0,\tau)=\left|\sum_l a_{m,l}(q;t_{f}^\prime, t_{i}^\prime)a_{l,m}(q;t_f,t_i)\right|^2$, which can be reduced to
\begin{equation}  \label{eq:2plsgeneral}
    P_m=\left|\sum_l |a_{l,m}|^2 e^{i\Phi_{l,m}}\right|^2,
\end{equation}
with the interference phase $\Phi_{l,m}=E_l(q)\tau/\hbar+2\phi_{l,m}$. There are two components in the interference phase. Besides the kinetic phase $E_l(q)\tau/\hbar$ of the intermediate state $|2l\hbar k+q\rangle$ during the interval $\tau$, the diffraction phase $\phi_{l,m}$ enters the interference, with a factor $2$. This factor $2$ comes from the two KD pulses, with each one contributing a phase $\phi_{l,m} (\phi_{m,l})$. As $a_{l,m}=a_{m,l}$, the phases from the two pulses add up rather than cancel each other. It is this term $2\phi_{l,m}$ that we can observe the effect of the diffraction phases. Generally speaking, the phase $\phi_{l,m}$ of different $l$ does not equal so that the overall interference phase $\Phi_{l,m}$ is shift by a different amount, resulting in a deformation in the dependence of $P_m$ on $\tau$.

Because of its ultra-narrow momentum spread (the root mean square width of a typical $^{87}$Rb BEC with atom number $1\times10^5$ is $\Delta p_{rms}<0.01\hbar k$), the ground state of a BEC can be approximated by $|p=0\rangle$, which means $m=0$ and $q=0$. Additionally, as $s=20$, the highest significantly populated diffraction order $n_{max}=\lfloor\sqrt{s} /2\rfloor=2$~\cite{Meystre2001book, Ovchinnikov1999PRL}, so $l$ can be limited to $-2\leq l\leq 2$. Equation~(\ref{eq:2plsgeneral}) can be approximated well as
\begin{equation}
    \begin{split}
        P_{0}=& C_{0}-C_{1}\cos (t+\varphi _{0,1})-C_{3}\cos (3t+\varphi _{1,2}) \\
        & +C_{4}\cos (4t+\varphi _{0,2}),
    \end{split}
    \label{eq:2pls}
\end{equation}
where $t=4\omega _{R}\tau $ and $\varphi _{m,l}=2(\phi _{m}-\phi_{l})-(m-l)\pi $. Here  $C_{0}=|a_{0}|^{4}+4|a_{1}|^{4} +4|a_{2}|^{4}$ is an offset, and $C_{i}(i=1,3,4)$ is the oscillation amplitude of each frequency component, respectively, with $C_1=4|a_0|^2|a_1|^2$, $C_3=8|a_1|^2|a_2|^2$ and $C_4=4|a_0|^2|a_2|^2$.
In most experiments realized in the RN regime, $a_2\ll a_1\thicksim a_0$, hence $C_3$ and $C_4$ are much smaller than $C_1$, which makes it reasonable to neglect these multi-frequency terms. However, in our case, with the experimental parameters described above, we have $|a_l|\simeq J_l(\Theta)$ with $\Theta\simeq1.2$, which can be seen from Fig.~\ref{fig:phase}. This means that $|a_0|$ ,$|a_1|$ and $|a_2|$ are of the same order, while the former two are only several times larger than the latter one. In this case, $C_1$ is only about an order of magnitude larger than $C_3$ and $C_4$, which means that the multi-frequency terms cannot be neglected any more.
%

From Eqs.~(\ref{eq:2plsgeneral}) and (\ref{eq:2pls}) we can see that the oscillation of $P_0$ with $\tau$ comes from the cosine functions whose frequencies are multiples of the recoil frequency. The oscillation period is then the Talbot period~\cite{Deng1999PRL, Cronin2009rmp} $T_T=h/4E_R \approx 79~\mu$s. The periodicity can be understood as follows. During the interval, atoms in each momentum state evolve freely according to their kinetic energy, accumulating phases from $0$ to $2\pi$. As the phases change with the interval, the probability amplitudes of atoms in different momentum states interfere constructively and destructively periodically when combined by the second pulse~\cite{Edwards2010pra, Xiong2011pra}.
%

The effect of the phase difference in each cosine function is as follows. According to Eq.~(\ref{eq:2pls}), there are three frequency components in the oscillation of $P_0$ with $\tau$. The amplitude ratio of each frequency component determines the number of peaks in one period of the Talbot time. As we have argued, $C_1$ is about an order of magnitude larger than $C_3$ and $C_4$. The dependence of $P_0$ on $\tau$ is then determined by the biggest amplitude single-frequency term $\cos(t)$, giving only one peak over a Talbot period. The phase difference $\varphi_{10}$ in the single-frequency term shifts the peak by $\Delta \tau$. The other two multi-frequency terms add small distortion to the peak produced by the single-frequency term. Shoulders appear when the distortion is strong enough to be comparable with the amplitude of the single-frequency term, but not so strong to produce additional peaks. The shift of the maximum position by the diffraction phase difference $\varphi_{0,1}$ results in distortions appearing at different distance to the maximum but with same strength. The inclusion of the diffraction phase differences $\varphi_{0,2}$ and $\varphi_{1,2}$ further enhances the asymmetry. The collaboration of the phase differences results in different distortion strength as what we have observed in the experiments. Numerical result according to Eq.~(\ref{eq:2pls}) is shown by the red solid line in Fig.~\ref{fig:mainresult}. We see that its agreement with the experimental data is good (the damping will be discussed later).

Without considering the phase factor $\phi_{m,l}$ in each term, Eq.~(\ref{eq:2pls}) is reduced to RN approximation, which is described in details in~\cite{Edwards2010pra}. In Fig.~\ref{fig:mainresult}, the result by RN approximation (oliver dotted line) is also shown for comparison. We can see that the peak of RN approximation locates exactly at $\tau=(2n+1)T_T/2 (n=0, 1, 2)$. The shift of the peak by the phase difference $\varphi_{10}$ is $\Delta \tau=\varphi_{10}/2\pi \times T_T$. In our case, $\varphi_{10}\simeq 0.16\pi$ gives $\Delta\tau\simeq 6.3~\mu$s, which agrees with the experimental result of $\Delta \tau=6.5~\mu$s (see the detail in Fig.~\ref{fig:mainresult}(b), the maximum of experimental data lies at $\tau=33~\mu$s while that of RN approximation lies exactly at half the Talbot period $39.5~\mu$s). Besides, the result by RN approximation is symmetric in a period, which further confirms our argument that the asymmetry is induced by the diffraction phase differences.

We notice that in~\cite{Buchner2003pra}, B\"uchner \emph{et al.} also study the phase shift of the diffraction process in the contrast interferometer built by Gupta \emph{et al.}~\cite{Gupta2002prl}. However, they ignore the $\pm4\hbar k$ term, which makes the final oscillation monochromatic (see Eq.~(15) in~\cite{Buchner2003pra}). Although a phase difference between the zeroth and first order diffractions is introduced, the monochromatic oscillation will not give rise to any asymmetry except a trival shift.

As the interval increases, the interaction between atoms in the condensate shows its effect. From Fig.~\ref{fig:mainresult}(a) we can see that, as the oscillation period increases, the theory without considering the interatomic interactions deviates from the experimental data. The deviation comes from the interatomic interaction, which can be described by the nonlinear term in the one-dimensional Gross-Pitaevskii (GP) equation~\cite{Trippenbach2000jpb, Isella2005pra}
\begin{equation}
\begin{split}
i\hbar\frac{\partial}{\partial t}\psi(z) =&-\frac{\hbar^2}{2M} \frac{\partial^2} {\partial z^2}\psi(z) +f(t)V(z)\psi(z)\\ &+g_{1d}|\psi(z)|^2 \psi(z).
\end{split}
\end{equation}
Here the condensate wavefunction $\psi(z)$ is normalized to the total atom number $N_0$ and $g_{1d}=2\hbar\omega_\bot a_s$ is the effective one-dimensional interaction strength~\cite{Isella2005pra}, with $a_s$ the $s$-wave scattering length, and $\omega_\bot$ the harmonic oscillation frequency in the radial direction. Using mature numerical methods~\cite{Bao2003jcp, Muruganandam2009cpc}, we numerically compute the ground state of a BEC in a harmonic trap and solve the time-dependent GP equation. Substituting the numerically obtained ground state into the time-dependent GP equation, we can get the relative population of the zero momentum state after the two pulses. With the experimental parameters described previously, the simulation results of different interval $\tau$ are shown by the blue dashed line in Fig.~\ref{fig:mainresult}. We can see that its consistency with the experimental data is excellent, even when the interval is as large as $\tau=300\thicksim400~\mu$s.


It is pointed out that, the main effect of the interatomic
interaction is to change the momentum distribution of the ground
state of the BEC in the harmonic trap, which leads to the
oscillation damping in Fig.~\ref{fig:mainresult}(a). In our
simulation, when the interatomic interaction is changed, the
momentum spread of the ground state of BEC changes accordingly,
leading to different damping rate. However, keeping the same ground
state, if we change the interatomic interaction strength during the
TL interferometer process, the damping rate changes little. The
oscillation damping comes from the momentum distribution of the
ground state also can be reflected by the compare with the case
without the atomic interaction. In Fig.~\ref{fig:mainresult}, the
initial state $|p=0\rangle$ is replaced by a Gaussian wave packet with a root-mean-square momentum width $\Delta p=0.02\hbar k$, then the damping rates from the results of Eq.~(\ref{eq:2pls}) and RN approximation are nearly the same as the experimental one. The interaction absent damping apparently comes from the initial finite momentum spread. The slightly larger value of the momentum spread
may be caused by additional damping mechanics such as the ignored
heating effect of the BEC. Hagley \emph{et al.}~\cite{Hagley1999prl}
and Campbell \emph{et al.}~\cite{Campbell2005prl} explained the
damping as the finite coherent length of the BEC. According to
Heisenberg's uncertainty principle, a finite coherent length
corresponds to a finite momentum spread, hence that is consistent
with our explanation.

\section{Conclusion}

In conclusion, we have studied the diffraction of an atom from a standing-wave pulse over a wide regime via the evolution operator method. The advantage of this method is that it is not limited by the pulse intensity and duration. Approximate analytical expressions for the diffraction amplitudes and diffraction phases of different orders are obtained. The effect of the diffraction phases is further studied experimentally using a temporal TL interferometer with a BEC. An asymmetry between the rising and falling edges is observed in the population oscillation. The experimental result agrees well with our analysis including the high-order diffraction phases. Our work implies that the diffraction phase plays an essential role in getting a better quantitative analysis of the matter wave interference experiments, particularly in precision measurements.

\section*{Acknowledgements}

We thank Peng Zhang, Biao Wu, Hui Zhai and Han Pu for helpful discussions. This work is supported by NKBRSFC (2011CB921501),NSFC (61027016,10934010,11175246),the RFDP number 20120001110091 and NCET-10-0209.

\end{CJK}
\end{document}